\documentclass[pre,showpacs,amsmath,amssymb]{revtex4}
\usepackage{graphicx}% Include figure files
\usepackage{dcolumn}% Align table columns on decimal point
\usepackage{bm}% bold math
\begin{document}
\title{Synchronous Dynamics of a Hopfield Model with Random Asymmetric Interactions}
\author{Manoranjan P. Singh} 
\affiliation{Laser Physics Division, Centre for Advanced Technology, Indore 452 013, India}
\email{mpsingh@cat.ernet.in}
\date{7 July 2003}

\begin{abstract}
We study the synchronous dynamics of the Hopfield model when a random antisymmetric part is added to the otherwise symmetric synaptic matrix. We use a generating functional technique to derive analytical expressions for the order parameters at the first time step ($t=1$) and the second time step ($t=2$). We find that the overlap between the target pattern and the state of the network at $t=1$ is independent of the symmetry of the additional random part of the synaptic interaction matrix. The result may have bearing on the theories which use the results at $t=1$ to estimate quantities relevant to the retrieval performance of the network. The symmetry of the synaptic interaction matrix becomes effective from the second time step, explicitly through a correlation function involving spin configurations at different times. This suggests that the prediction of the long time behavior of the network from the first few time steps may not be {\it always} correct. This was confirmed by the numerical simulation which shows that the difference in the long time and the short time behavior of the network becomes pronounced in presence of asymmetry in the synaptic interaction matrix. It is also found in  simulations that the size of the basins of attraction of the stored patterns decreases with an increase in the asymmetry. Moreover, the convergence time for the retrieval  increases. These results are contrary to the expectations from the earlier studies based on the counting of fixed points. However, the convergence time for the spurious fixed points increases  faster than that for the retrieval fixed points in the presence of asymmetry in the synapses. This is a positive feature of the asymmetry so far as the retrieval performance of the network is concerned. 
\end{abstract}
\maketitle
\section{Introduction}
Theoretical analyzes of neural network models of associative memory using the techniques of statistical physics, particulary the ones pertaining to the area of spin-glasses  have  contributed enormously to the understanding of these models \cite{amit,parisibook,hkp}. The most  widely analyzed
model of this kind is the one proposed by 
Hopfield \cite{hop1,hop2}. In the Hopfield
model, a set of random binary patterns (``memories'') are stored in a
network using the modified Hebb rule which leads to a symmetric
matrix of synaptic interactions. While the  assumption
of symmetry of the synaptic interaction matrix enables us to use the well developed analytical tools of the equilibrium statistical mechanics, it is known \cite{eccles}
to be biologically unrealistic.  

Furthermore, studies on models of spin glasses and neural networks suggest that presence of asymmetry of an appropriate magnitude in the synaptic interaction matrix may result in the improvement of the performance of the network as an associative memory (see Refs. \cite{us1, us2} for a detailed discussion on this aspect). It is well-known
\cite{amit} that the Hopfield model has a large number of
spurious fixed-point attractors which are not strongly correlated
with any one of the memories stored in the network. The presence of these
attractors adversely affects the performance of the network by
``trapping'' initial configurations which are not close to one of
the memory states. This problem is aggravated by the fact that the
network can not, in any way, distinguish between successful retrieval of
a memory and convergence to a spurious fixed point because in both
cases it settles down to a time-independent state. The presence of
asymmetry in the interactions is believed to alleviate this problem
in two different ways. First, analytic and numerical studies
\cite{us1,Tr88,fuk90} have shown that the introduction of asymmetry greatly
reduces the number of spurious fixed-point attractors while leaving
the attractors corresponding to memory states largely unaffected.
These results suggest that the basins of attraction of the memory
states may be enlarged by the introduction of asymmetry. Second, there
are indications \cite{parisi} that the presence of asymmetry
greatly increases the time of convergence to spurious attractors
relative to that for convergence to memory states. Thus it may be possible to discriminate between the actual retrieval of a stored pattern and a spurious retrieval through the dynamical behavior of the network which has an asymmetry in the synaptic matrix. This has been confirmed in the 
numerical simulations on the Hopfield model with sequential dynamics \cite{us2}. It is also worth mentioning here that the non-equilibrium dynamics of such models can exhibit the  features  similar to those observed in the glassy systems, which constitute an area of current research \cite{glassyreview}. 

In this paper, we describe the results of an analytical study which may be relevant 
to the understanding of 
retrieval properties of a Hopfield-like model in which a  random 
antisymmetric component is added to the symmetric, Hebbian synaptic matrix. In a sense this 
study complements our earlier studies on the same model \cite{us1, us2}. Further, it may be useful 
in developing an analytical technique to study the dynamics of the model, which remains an open problem. The major challenge before a reasonable description of the dynamical behavior is  to 
properly account for the noise term in the local field arising due to the extensive loading of the memory. The presence of asymmetry in the synaptic interaction matrix makes the task further complicated.

In some of the simple phenomenological theories, for example, in Refs. \cite{nishimori93, shukla93}, the noise term is assumed to have a Gaussian distribution. However, this assumption is shown to have very limited validity \cite{nishimori93,henkelOpper91,coolen93,coolen94}. Although a dynamical replica theory  \cite{coolen93, coolen94} has been used to study the dynamics of the Hopfield model at finite time scales, it is not clear whether it can be extended to the asymmetric Hopfield model considered here. 

We have used a generating functional technique to analyze the synchronous dynamics of the asymmetric Hopfield model. This technique  has been applied to study the synchronous dynamics of the (symmetric) Hopfield model \cite{gardner87}. It is exact in the thermodynamic limit. The choice of  the synchronous dynamics is because of its amenability to the generating functional method. However, most of the results are expected to hold true for the asynchronous dynamics too.  Although the generating functional formalism can be used, at least  in principle, to calculate the properties of the network after an arbitrary number of time steps. There is a major problem in implementing the procedure for larger time steps --  the number of order parameters increase very quickly with time. We, therefore, restrict ourselves here to the analytic expressions for the overlap with a stored pattern after one and two time steps. 

We find that the overlap (between the target pattern and the state of the network) after one time step to be independent of the symmetry of the additional random part of the synaptic interaction matrix. So far as the overlap after one time step is concerned the additional random part of the synaptic matrix  acts just like a noise  which goes with the noise term which results from the extensive loading of the memory. Symmetry of the synaptic matrix, however, comes into picture when we calculate the overlap after two time steps. In this the information of  the symmetry of the  additional random part of the synaptic matrix comes explicitly through an order parameter which involves the correlation between the neuron configurations at different times. Similar result has also been reported \cite{krauthNadalMezard88} for a one-pattern model of associative memory. This feature makes the dynamics of the asymmetric Hopfield model quite different from that of the usual Hopfield model. In the usual Hopfield  model dynamics is often described approximately in terms of the recurrence relations based on the expression of the overlap after single time step. This has also been used to compute successfully the basin of attraction, the rate of convergence and the amount of error-correction of the model \cite{Ke88,kp93}. Obviously, such  approximate schemes will not generally hold good for the asymmetric Hopfield model.     

We have  checked the results of the analytical calculation by carrying out numerical simulation for a network of 500 neurons. We find that the result of the analytical calculation for the overlap after one and two time steps, $m_1^1$ and $m_1^2$, agrees very well with that obtained by the numerical simulation. Values of both $m_1^1$ and $m_1^2$ decrease with an increase in the asymmetry in the synapses. This is true for the value of overlap at large times too. Thus the retrieval quality undergoes degradation with the asymmetry. This is consistent with the conclusion drawn from the properties of the fixed point attractors in Ref. \cite{us1}. We also find that the overlap $m_1^t$ is not always a monotonic function of time. For smaller values of the initial overlap $m_1^0$, $m_1^t$ increases with time for first few time steps and decreases thereafter to converge  to its asymptotic limit $m_1^{\infty}$. The similar behavior has  been reported in Ref. \cite{gardner87} for the synchronous dynamics of the Hopfield model. However, the introduction of asymmetry makes the difference in the long time and the short time behavior of the network much more pronounced. We have also calculated in the simulation the probability of convergence of the network to a state corresponding to the target memory and that to a spurious attractor. The corresponding average convergence times were also calculated. We find that the probability of convergence to the retrieval fixed point decreases with asymmetry while that to the spurious fixed point increases. However, as reported for the asynchronous dynamics of the same model \cite{us2}, the convergence time for the spurious fixed points increases  faster than that for the retrieval fixed point in presence of asymmetry in the synapses. This behavior may be quite useful in providing a way to discriminate the correct retrieval from a spurious one. 

The paper is organized as follows. In Section \ref{generatingfunctional} we describe the generating functional technique and obtain  the analytic expressions for the order parameters after one and two time steps. We discuss the results in Section \ref{results}. The paper is concluded in Section \ref{summary}. The technical details of the derivations are presented in Appendices \ref{appen1} and \ref{appen2}. 
\section{Generating Functional Method}
\label{generatingfunctional}
We consider a system of $N$ neurons (``spins") $\{\sigma_i\}$, each of which can take values $+ 1$ or $-1$. The synaptic interaction matrix, which describes how neuron $i$ affects the state of neuron $j$, is defined as 
\begin{eqnarray}
J_{ij}&=&J_{ij}^{\rm s}+k J_{ij}^{\rm as}\qquad \text{for} \qquad i \ne j \nonumber \\
&=& 0 \qquad \text{for} \qquad i=j\,.
\end{eqnarray}
$J_{ij}^{\rm s}$ which describes the symmetric part of the synaptic matrix is given by the modified Hebb's rule in terms of the stored patterns $\{\xi_i^{\mu}\}$, $\mu=1,\ldots,p$
\begin{equation}
J_{ij}^{\rm s}=\frac{1}{N} \sum_{\mu=1}^{p} \xi_i^{\mu}\xi_j^{\mu}\,.
\end{equation}
Here, each $\xi_i^{\mu}$ may take values $\pm 1$ with equal probability. The number of stored patterns in the network is $p$, and $\alpha=p/N$ is the memory loading of the network. The elements of the  antisymmetric part $J_{ij}^{\rm as}$ of the synaptic matrix are drawn randomly from a Gaussian distribution with zero mean and variance $1/\sqrt{N}$, that is,
\begin{equation}
P(J_{ij}^{\rm as})=\frac{1}{\sqrt{2\pi/N}}\exp\left[{\frac{-(J_{ij}^{\rm as})^{2}}
{2/N}}\right], \,\, i<j ,
%\label{eqn3}
\end{equation}
and $ J_{ij}^{\rm as}=-J_{ji}^{\rm as}$. The parameter $k$ is a measure of the strength
of the antisymmetric part of the coupling relative to that of the
symmetric part. The synchronous dynamics of the system is given by the rule
\begin{equation}
\sigma_i^t=\mathrm{sgn}\left(\sum_{i=1}^N J_{ij}\sigma_j^{t-1}\right)\,,
\end{equation}
where all the spins are updated at the same time. The dynamics of the system can be described by the following generating functional \cite{gardner87} for the average spin-spin correlation function
\begin{equation}
y(h_{t_1 t_2}) = \left\langle \mathrm{Tr}^{\prime}_{\{\sigma_i^0\}}
\underset{t = 1,\ldots,T}{\mathrm{Tr}_{\{\sigma_i^t\}}}
\left[ \prod^N_{i=1} \prod^T_{t=1} \Theta \left( \sigma_i^t 
\sum_{j=1}^N J_{ij} \sigma_j^{t-1} \right) \exp\left( \sum_{t_1t_2} 
h_{t_1t_2} \sum_t \sigma_i^{t_1} \sigma_{i}^{t_2} \right) \right] 
\right \rangle_{\{J_{ij}\}},
\end{equation}
where $\langle \cdots \rangle_{\{J_{ij}\}}$ indicates and average 
over the realizations of the synaptic interaction matrix $\{J_{ij}\}$. 
The trace, Tr$^{\prime}_{\{\sigma_i^0\}}$ over the initial spin 
configuration is restricted to 
\begin{equation}
\sigma_i^0=
\begin{cases}
-\xi^1_i&  \quad 1\,\le\, i\, \le\, Ng,\\
\xi^1_i& \quad Ng\,<\,i\, \le\,N,
\end{cases}
\label{config0}
\end{equation}
so that it has overlap $1-2 g$ with the $\mu = 1$ pattern. Using the integral 
representation for  the $\theta$-function, $y$ can be expressed as 
\begin{eqnarray}
y(h_{t_1 t_2}) &= &\left\langle \mathrm{Tr}^{\prime}_{\{\sigma_i^0\}}
\,{\mathrm{Tr}_{\{\sigma_i^t\}}}
\exp\left( \sum_{t_1t_2} 
h_{t_1t_2} \sum_t \sigma_i^{t_1} \sigma_{i}^{t_2} \right) 
\prod_{i,t} \left[ \int_0^{\infty} d\lambda_i^t \int_{-\infty}^{\infty} 
\frac{d x_i^t}{2\,\pi} \right] \times \right. \nonumber \\
&&\left. \exp\left(i \sum_{i,t} \lambda_i^t x_i^t -i \sum_{i,j,t}  
x_i^t \, \sigma_i^t \, J_{ij}\, \sigma_j^{t-1} \right) 
\right\rangle_{\{J_{ij}\}}.
\end{eqnarray}
For the synaptic matrix considered here  we have 
\begin{equation}
\left\langle \exp\left(-i \sum_{i,j,t}  
x_i^t \, \sigma_i^t \, J_{ij}\, \sigma_j^{t-1} \right) 
\right\rangle_{\{J_{ij}\}} = X \,\, Y,
\end{equation}
where $X$ and $Y$  contain averages over the symmetric part (stored patterns) 
and the antisymmetric part of the synaptic interaction matrix, respectively. These are 
given by
\begin{eqnarray}
&&X = \left\langle \exp \left(-\frac{i}{N} \sum_{\mu=1}^p \underset{i \ne j} 
{\sum_{i,j}} x_i^t \sigma_i^t \xi_i^{\mu} \xi_j^{\mu} \sigma_j^{t-1} \right) \right\rangle_{\{\xi_i^{\mu}\}} \nonumber \\
&& = \left\langle \exp \left(-\frac{i}{N} \sum_{\mu,t} 
\left\{\sum_i x_i^t \sigma_i^t \xi_i^{\mu} \right\} 
\left\{ \sum_i \xi_i^{\mu} \sigma_i^{t-1}\right\} + i \alpha \sum_{i,t} x_i^t
\sigma_i^t \sigma_i^{t-1} \right) \right\rangle_{\{\xi_i^{\mu}\}}
\end{eqnarray}
\begin{equation}
Y = \left\langle \exp \left( -i \sum_{i,j,t} J_{ij}^{\rm as} x_i^t \sigma_i^t 
\sigma_j^{t-1} \right) \right\rangle_{\{J_{ij}^{\rm as}\}}.
\end{equation}

In order to carry out the average over the stored patterns it is convenient 
to define  variables
\begin{equation}
m_{\mu}^{t-1}=\frac{1}{N}\sum_i \xi_i^{\mu} \sigma_i^{t-1},
\end{equation}
and their conjugates  $n_{\mu}^{t-1}$ through the identity
\begin{equation}
1 \equiv \prod_{\mu,t} \left[ \int_{-\infty}^{\infty} N\, dm_{\mu}^{t-1} 
\int_{-\infty}^{\infty} \frac{dn_{\mu}^{t-1}}{2\,\pi} \exp\left( i 
N n_{\mu}^{t-1} m_{\mu}^{t-1} - i n_{\mu}^{t-1} \sum_i \sigma_i^{t-1} 
\xi_i^{\mu} \right) \right],
\end{equation}
so that $X$ becomes,
\begin{eqnarray}
X&=& \int \left[\frac{N\, dm_{\mu}^{t-1}\,dn_{\mu}^{t-1}}{2\,\pi} 
\right] \exp \left( i N \sum_{\mu,t} n_{\mu}^{t-1} m_{\mu}^{t-1} 
+i \alpha \sum_{i,t} x_i^t \sigma_i^t \sigma_i^{t-1} \right)\nonumber \\ 
&&\left\langle \exp \left( - i \sum_{\mu,t}n_{\mu}^{t-1} 
\sum_i \sigma_i^{t-1} \xi_i^{\mu} -
i \sum_{\mu,t} m_{\mu}^{t-1} \sum_i x_i^t \sigma_i^t \xi_i^{\mu} \right) 
\right\rangle_{\{\xi_i^{\mu}\}}\, .
\end{eqnarray}
After averaging over the patterns, $\mu =2,\ldots,p$ we get
\begin{eqnarray}
X&=& \int \left[\frac{N\, dm_{\mu}^{t-1}\,dn_{\mu}^{t-1}}{2\,\pi} 
\right] \exp \left( i N \sum_{\mu,t} n_{\mu}^{t-1} m_{\mu}^{t-1} 
+i \alpha \sum_{i,t} x_i^t \sigma_i^t \sigma_i^{t-1} \right)\nonumber \\ 
&&\exp\left( \sum_{\mu=2}^p \sum_i \ln\left\{\cos\left[\sum_{t=1}^T
\left(n_{\mu}^{t-1} \sigma_i^{t-1} + m_{\mu}^{t-1} x_i^t \sigma_i^t \right)\right]\right\}
\right) \nonumber \\
&&\left\langle  \exp \left( - i \sum_t n_1^{t-1} 
\sum_i \sigma_i^{t-1} \xi_i^1 -
i \sum_t m_1^{t-1} \sum_i x_i^t \sigma_i^t \xi_i^1 \right)
\right\rangle_{\{\xi_i^1\}}\, .
\end{eqnarray}

Since the spin configuration has a finite overlap only with  pattern 1, only 
the first term in the expansion of $\ln \cos[\quad]$ contributes to leading order in 
$N$. Thus we get
\begin{eqnarray}
X&=&\int \left[\frac{N\, dm_{\mu}^{t-1}\,dn_{\mu}^{t-1}}{2\,\pi} 
\right] \exp \left( i N \sum_{\mu,t} n_{\mu}^{t-1} m_{\mu}^{t-1} 
+i \alpha \sum_{i,t} x_i^t \sigma_i^t \sigma_i^{t-1} \right)\nonumber \\ 
&& \exp \left(-\frac{1}{2} \sum_{t_1,t_2} \left[\sum_{\mu > 1} n_{\mu}^{t_1-1}
n_{\mu}^{t_2-1} \sum_i \sigma_i^{t_1-1}\sigma_i^{t_2-1} + \sum_{\mu > 1} 
m_{\mu}^{t_1-1} m_{\mu}^{t_2-1} \sum_i x_i^{t_1} x_i^{t_2} \sigma_i^{t_1} 
\sigma_i^{t_2} \right. \right. \nonumber \\
&&\left. \left.+ 2 \sum_{\mu > 1} n_{\mu}^{t_2-1} m_{\mu}^{t_1-1} 
\sum_i x_i^{t_1} \sigma_i^{t_1} \sigma_i^{t_2-1}\right]
\right) \nonumber \\
&&\left\langle  \exp \left( - i \sum_t n_1^{t-1} 
\sum_i \sigma_i^{t-1} \xi_i^1 -
i \sum_t m_1^{t-1} \sum_i x_i^t \sigma_i^t \xi_i^1 \right)
\right\rangle_{\{\xi_i^1\}}\,.
\end{eqnarray}

Now
\begin{eqnarray}
Y= \underset{i < j}{\prod_{i,j}} \left[ \int_{-\infty}^{\infty}
\frac{dJ_{ij}^{\rm as}}{\sqrt{2\,\pi/N}} \exp \left\{ \frac{N\,\left( J_{ij}^{\rm as}\right)^2}{2} - i k J_{ij}^{\rm as} 
\sum_t \left(x_i^t \sigma_i^t \sigma_j^{t-1} - x_j^t \sigma_j^t 
\sigma_i^{t-1} \right) \right\} \right] \nonumber \\
=\prod_{i,j} \exp \left\{- \frac{k^2}{2 N} \sum_{t_1,t_2} 
\left(x_i^{t_1} \sigma_i^{t_1} \sigma_j^{t_1-1} - x_j^{t_1} \sigma_j^{t_1} 
\sigma_i^{t_1-1} \right) \left(x_i^{t_2} \sigma_i^{t_2}\sigma_j^{t_2-1} 
- x_j^{t_2} \sigma_j^{t_2} \sigma_i^{t_2-1}\right)\right\}.
\end{eqnarray}
Ignoring terms of ${\mathcal{O}}(1/N)$ we get 
\begin{equation}
Y = \exp\left\{\frac{k^2}{2\,N} \sum_{t_1,t_2}\left(\sum_i x_i^{t_1}
\sigma_i^{t_1}\sigma_i^{t_2-1} \sum_i x_i^{t_2}\sigma_i^{t_2}\sigma_i^{t_1-1}
- \sum_i x_i^{t_1} x_i^{t_2} \sigma_i^{t_1}\sigma_i^{t_2}
\sum_i \sigma_i^{t_1 - 1}\sigma_i^{t_2-1} \right) \right\}\,.
\end{equation}

Let
\begin{equation}
\frac{1}{N} \sum_i \sigma_i^{t_1 - 1} \sigma_i^{t_2-1}=q^{t_1 t_2} \quad \text{for} \quad t_1 < t_2,
\end{equation}
\begin{equation}
\frac{1}{N}  \sum_i x_i^{t_1} \sigma_i^{t_1} \sigma_i^{t_2-1} = s^{t_1 t_2},
\end{equation}
\begin{equation}
\frac{1}{N} \sum_i x_i^{t_1} x_i^{t_2} \sigma_i^{t_1} \sigma_i^{t_2} = p^{t_1 t_2} \quad \text{for} \quad t_1 \le t_2\,.
\end{equation}
Introducing the delta function constraints corresponding to the variables defined 
above:
\begin{eqnarray}
1 = \underset{t_1 < t_2} {\prod_{t_1,t_2}} \left[ \int_{-\infty}^{\infty} N\, dr^{t_1 t_2} 
\int_{-\infty}^{\infty} \frac{dq^{t_1 t_2}}{2\,\pi} \exp\left( i 
N r^{t_1 t_2} q^{t_1 t_2} - i r^{t_1 t_2} \sum_i \sigma_i^{t^1-1} 
\sigma_i^{t^2-1} \right) \right], \\
1 =  \prod_{t_1,t_2} \left[ \int_{-\infty}^{\infty} N\, dk^{t_1 t_2} 
\int_{-\infty}^{\infty} \frac{ds^{t_1 t_2}}{2\,\pi} \exp\left( i 
N k^{t_1 t_2} s^{t_1 t_2} - i k^{t_1 t_2} \sum_i x_i^{t^1}\sigma_i^{t^1} 
\sigma_i^{t^2-1} \right) \right], \\
1 =  \underset{t_1 \le t_2} {\prod_{t_1,t_2}} \left[ \int_{-\infty}^{\infty} N\, dl^{t_1 t_2} 
\int_{-\infty}^{\infty} \frac{dp^{t_1 t_2}}{2\,\pi} \exp\left( i 
N l^{t_1 t_2} p^{t_1 t_2} - i l^{t_1 t_2} \sum_i x_i^{t^1} x_i^{t^2} \sigma_i^{t^1} \sigma_i^{t^2} \right) \right], 
\end{eqnarray}
and performing average over $\{\xi_i^1\}$ in accordance with the constraint 
given by Eq. (\ref{config0}) we get
\begin{equation}
y\left(h_{t_1 t_2}\right) \propto \int_{\infty}^{\infty} dm_1^{t-1}\, dn_1^{t-1}\, 
dr^{t_1 t_2}\, dq^{t_1 t_2}\, dk^{t_1 t_2}\, ds^{t_1 t_2}\, dl^{t_1 t_2}\, 
dp^{t_1 t_2}\, \mathrm{e}^{N\,F\left(m_1, n_1, \ldots, h_{t_1 t_2}\right)},
\label{sdintegration}
\end{equation} 
where
\begin{eqnarray}
F\left(m_1, n_1, \ldots, h_{t_1 t_2}\right)&=& -\frac{k^2}{2} \sum_t p^{t\,t} 
- k^2 \sum_{t_1 < t_2} p^{t_1\,t_2}\,q^{t_1\,t_2} + \frac{k^2}{2}\sum_{t_1 t_2}
s^{t_1\,t_2} s^{t_2\,t_1} \nonumber \\
&& +\,\, i \sum_{t_1 < t_2} r^{t_1\,t_2}\,q^{t_1\,t_2}+i \sum_{t_1 \le t_2}
l^{t_1\,t_2}\, p^{t_1\,t_2} + i \sum_{t_1 t_2} k^{t_1\,t_2}\,s^{t_1\,t_2} \nonumber \\
&& +\,\, i \sum_i n_1^{t-1}\,m_1^{t-1} + \alpha \,\ln{W}\left(q^{t_1\,t_2},\,
s^{t_1\,t_2},\,p^{t_1\,t_2}\right) \nonumber \\
&&+\,\,(1-g)\,\ln{Z_+}\left(n_1^t,\,m_1^t,\,r^{t_1\,t_2},\,k^{t_1\,t_2},\,l^{t_1\,t_2},\,
h_{t_1\,t_2}\right) \nonumber \\
&&+\,\,g\,\ln{Z_-}\left(n_1^t,\,m_1^t,\,r^{t_1\,t_2},\,k^{t_1\,t_2},\,l^{t_1\,t_2},\,
h_{t_1\,t_2}\right)\,,
\end{eqnarray}
\begin{eqnarray}
W &=& \int_{\infty}^{\infty} \prod_{t=1}^{T} \left[ \frac{N\,dm^{t-1}\,dn^{t-1}}{2\,\pi}\right] \exp \left\{N \sum_t \left[i n^{t-1} m^{t-1} - \frac{1}{2} \left(n^{t-1}\right)^2 - \frac{1}{2} p^{t\,t} \left(m^{t-1}\right)^2 \right] \right. \nonumber \\
&& \left.  - N\sum_{t_1 < t_2}\left[ q^{t_1\,t_2}\,n^{t_1-1}\,n^{t_2-1} +p^{t_1\,t_2}\,m^{t_1-1}\,m^{t_2-1} 
+ s^{t_1\,t_2} m^{t_1-1} n^{t_2-1}\right] \right\}\,,
\end{eqnarray}
and
\begin{eqnarray}
&&Z_{\pm}\left(n_1^t,\,m_1^t,\,r^{t_1\,t_2},\,k^{t_1\,t_2},\,l^{t_1\,t_2},\,
h_{t_1\,t_2}\right)=\frac{1}{2} \, \underset{t=0,1,\ldots,T} {\rm{Tr}_{\sigma^t}}
\int_0^{\infty} \prod_{t=1}^{T} \left[d\lambda^t\right] \int_{\infty}^{\infty} 
\prod_{t=1}^T \left[\frac{dx^t}{2\,\pi}\right]  \nonumber \\
&&\exp\left\{i \sum_{t=1}^T x^t\,\lambda^t + i\,\alpha\sum_{t=1}^T 
x^t\,\sigma^t\,\sigma^{t-1} + \sum_{t_1\,t_2} h_{t_1\,t_2}\,\sigma^{t_1}\,\sigma^{t_2} \right. \nonumber \\
&&\left.-i \sum_{t_1 < t_2} r^{t_1\,t_2}\,\sigma^{t_1-1}\,\sigma^{t_2-1}
 -i \sum_{t_1\,t_2} k^{t_1\,t_2}\,x^{t_1}\,\sigma^{t_1}\,\sigma^{t_2-1} -i \sum_{t_1 < t_2} l^{t_1\,t_2}\, x^{t_1}\,x^{t_2}\, \sigma^{t_1}\,
\sigma^{t_2}\right. 
\nonumber \\
&&\left.  \mp i \,\sigma^0 \sum_{t=1}^T n_1^{t-1}\,\sigma^{t-1} 
\mp i\,\sigma^0 \sum_{t=1}^T m_1^{t-1}\, x_t\,\sigma^t \right\} \,. 
\label{zpm}
\end{eqnarray}

In the limit $N\rightarrow \infty$ the integration in Eq. (\ref{sdintegration}) 
can be evaluated by steepest descents. The saddle point equations are
\begin{eqnarray}
%\begin{equation}
q^{t_1\,t_2}&  \underset{t_1 < t_2}{=}  &(1-g)\,\left\langle \sigma^{t_1-1}\,
\sigma^{t_2-1} \right\rangle_{Z_+} + g\,\left\langle \sigma^{t_1-1}\,
\sigma^{t_2-1} \right\rangle_{Z_-}\,,\label{speq1}\\
%\end{equation}
%\begin{equation}
s^{t_1\,t_2}&  =  &(1-g)\,\left\langle x^{t_1}\,\sigma^{t_1}\,
\sigma^{t_2-1} \right\rangle_{Z_+} + g\,\left\langle x^{t_1}\, \sigma^{t_1}\,
\sigma^{t_2-1} \right\rangle_{Z_-}\,,\\
%\end{equation}
%\begin{equation}
p^{t_1\,t_2}&  \underset{t_1 \le t_2}{=}  &(1-g)\,\left\langle x^{t_1}\,
x^{t_2}\,\sigma^{t_1}\,\sigma^{t_2} \right\rangle_{Z_+} + 
g\,\left\langle x^{t_1}\,x^{t_2}\,\sigma^{t_1}\,\sigma^{t_2} \right\rangle_{Z_-}\,,\\
%\end{equation}
%\begin{equation}
n_1^t& =  &(1-g)\,\left\langle \sigma^0\,x^{t+1}\,
\sigma^{t+1} \right\rangle_{Z_+} - g\,\left\langle \sigma^0\,x^{t+1}\,
\sigma^{t+1} \right\rangle_{Z_-}\,,\\
%\end{equation}
%\begin{equation}
m_1^t& = &(1-g)\,\left\langle \sigma^0\,
\sigma^t \right\rangle_{Z_+} - g\,\left\langle \sigma^0\,
\sigma^t \right\rangle_{Z_-}\,,\\
%\end{equation}
%\begin{equation}
i\,r^{t_1\,t_2}& \underset{t_1 < t_2}{=} &k^2\,p^{t_1\,t_2} + N\,\alpha\,\left\langle n^{t_1-1}\,n^{t_2-1} \right\rangle_W \label{rt1t2}\,,\\
%\end{equation}
%\begin{equation}
i\,k^{t_1\,t_2}& = &-k^2\,s^{t_1\,t_2} + N\,\alpha\,\left\langle m^{t_1-1}\,n^{t_2-1} \right\rangle_W\,,\\
%\end{equation}
%\begin{equation}
i\,l^{t_1\,t_2}& \underset{t_1 < t_2}{=} &k^2\,q^{t_1\,t_2} + N\,\alpha\,\left\langle m^{t_1-1}\,m^{t_2-1} \right\rangle_W\,,\\
%\end{equation}
%\begin{equation}
i\,l^{t\,t}& = &\frac{k^2}{2} +\frac{ N\,\alpha}{2}\,\left\langle \left(m^{t-1}\right)^2 \right\rangle_W\,.
%\end{equation}
\label{speq9}
\end{eqnarray}
The averages  are calculated with $h_{t_1\,t_2}=0$.

As shown in Appendix \ref{appen1}, a number of order parameters vanish:
\begin{eqnarray}
s^{t_1\,t_2}&=&0 \quad \text{for} \quad t_1 \ge t_2, \nonumber \\
p^{t_1\,t_2}&=&0 \quad \text{for all} \quad t_1,\,t_2 \quad \text{i.e.} \quad t_1 \le t_2, \\
&&\text{(note that $p^{t_1\,t_2}$ has been defined for $t_1 \le t_2$ only)} \nonumber\\
n_1^t&=&0 \quad \text{for all} \quad t\,.\nonumber
\end{eqnarray}
In addition following correlation functions also vanish:
\begin{equation}
\left\langle n^{t_1-1}\,n^{t_2-1}\right\rangle_W = 0 \quad \text{for} \quad t_1 < t_2 \,,\nonumber 
\end{equation}
and
\begin{equation}
\left\langle m^{t_1-1}\,n^{t_2-1}\right\rangle_W = 0 \quad \text{for} \quad t_1 < t_2 \,.
\end{equation}
Using $p^{t_1 t_2}=0$ we get
\begin{equation}
r^{t_1 t_2}=0\,.
\end{equation}

In view of these results the non-zero saddle point parameters are given as
\begin{eqnarray}
%\begin{equation}
q^{t_1\,t_2}&  \underset{t_1 < t_2}{=}  &(1-g)\,\left\langle \sigma^{t_1-1}\,
\sigma^{t_2-1} \right\rangle_{Z_+} + g\,\left\langle \sigma^{t_1-1}\,
\sigma^{t_2-1} \right\rangle_{Z_-}\,, \label{nonzerosp1} \\
%\end{equation}
%\begin{equation}
s^{t_1\,t_2}& \underset{t_1 < t_2}{=}  &(1-g)\,\left\langle x^{t_1}\,\sigma^{t_1}\,
\sigma^{t_2-1} \right\rangle_{Z_+} + g\,\left\langle x^{t_1}\, \sigma^{t_1}\,
\sigma^{t_2-1} \right\rangle_{Z_-}\,,\\
%\end{equation}
%\begin{equation}
m_1^t& = &(1-g)\,\left\langle \sigma^0\,
\sigma^t \right\rangle_{Z_+} - g\,\left\langle \sigma^0\,
\sigma^t \right\rangle_{Z_-}\,, \\
%\end{equation}
%\begin{equation}
i\,k^{t\,t}& = & N\,\alpha\,\left\langle m^{t-1}\,n^{t-1} \right\rangle_W\\
i\,k^{t_1\,t_2}& \underset{t_1 > t_2}{=} &-k^2\,s^{t_2\,t_1} + N\,\alpha\,\left\langle m^{t_1-1}\,n^{t_2-1} \right\rangle_W\,, \\
%\end{equation}
%\begin{equation}
i\,l^{t_1\,t_2}& \underset{t_1 < t_2}{=} &k^2\,q^{t_1\,t_2} + N\,\alpha\,\left\langle m^{t_1-1}\,m^{t_2-1} \right\rangle_W\,, \\
%\end{equation}
%\begin{equation}
i\,l^{t\,t}& = &\frac{k^2}{2} +\frac{ N\,\alpha}{2}\,\left\langle \left(m^{t-1}\right)^2 \right\rangle_W\,. \label{nonzerosp7}
%\end{equation}
\end{eqnarray}
Some of the non-zero parameter can be easily interpreted in  physical terms. For example, $q^{t_1 t_2}$ is the overlap between spin configurations at time $t_1-1$ and $t_2-1$, $s^{t_1 t_2}$ is related to the overlap of local fields and spins \cite{gardner87} and $m_1^t$ is the overlap between pattern 1 and the spin configuration at time $t$. Information of symmetry of the additional random part of the synaptic interaction matrix is explicitly contained in the order parameter $k^{t_1 t_2}$ for $t_1 \ne t_2$. 
\section{Results}
\label{results}

We have evaluated  explicitly  the non-zero order parameters for first two time steps (i.e. $t=1$ and 2). Here we present only the final result. Technical details of the calculation are given  in Appendix \ref{appen2}. The order parameters for $t=1$ are 
\begin{eqnarray}
k^{1\,1}&=& \alpha, \nonumber \\
i\,l^{1\,1}&=& \frac{1}{2} \left(\alpha + k^2\right), \nonumber \\
m_1^1&=&\mathrm{erf}\left\{\frac{m_1^0}{\sqrt{2\,\left(\alpha + k^2\right)}}\right\}\,.
\end{eqnarray}
We note that the one step dynamics does not depend on the symmetry of the 
additional random term to the synaptic interaction matrix of the Hopfield model. 
The additional part acts just like a noise which goes with the noise term resulting from the extensive loading of the memory. 
The new non-zero order parameters after two time steps are
\begin{eqnarray}
q^{1\,2}&=&m_1^0\,m_1^1, \nonumber \\
s^{1\,2}&=& i \sqrt{\frac{2}{\pi\,\left(\alpha+k^2\right)}}\,
\exp\left\{-\frac{(m_1^0)^2}{2\,(\alpha+k^2)}\right\},\nonumber \\
i \,l^{2\,2}&=& \frac{1}{2}\left(\alpha+k^2\right) -\frac{\alpha}{2}
\left\{\left(s^{1\,2}\right)^2+2\,i\,q^{1\,2}\,s^{1\,2}\right\}, \nonumber \\
i \,l^{1\,2}&=& \left(\alpha+k^2\right)q^{1\,2} -i\,\alpha\,s^{1\,2}, \nonumber \\
k^{2\,2}&=&\alpha, \nonumber \\
k^{2\,1}&=&-i \left(\alpha-k^2\right) s^{1\,2}, \nonumber \\
m_1^2&=&\frac{\left(1+m_1^0\right)}{2} 
\mathrm{erf}\left(\frac{m_1^1+k^{2\,1}}{2\,\sqrt{i\,l^{2\,2}}}\right)
+ \frac{\left(1-m_1^0\right)}{2} 
\mathrm{erf}\left(\frac{m_1^1-k^{2\,1}}{2\,\sqrt{i\,l^{2\,2}}}\right)\,.
\end{eqnarray}

\begin{figure}[htb]
\includegraphics[height=6.25cm,width=0.49\textwidth]{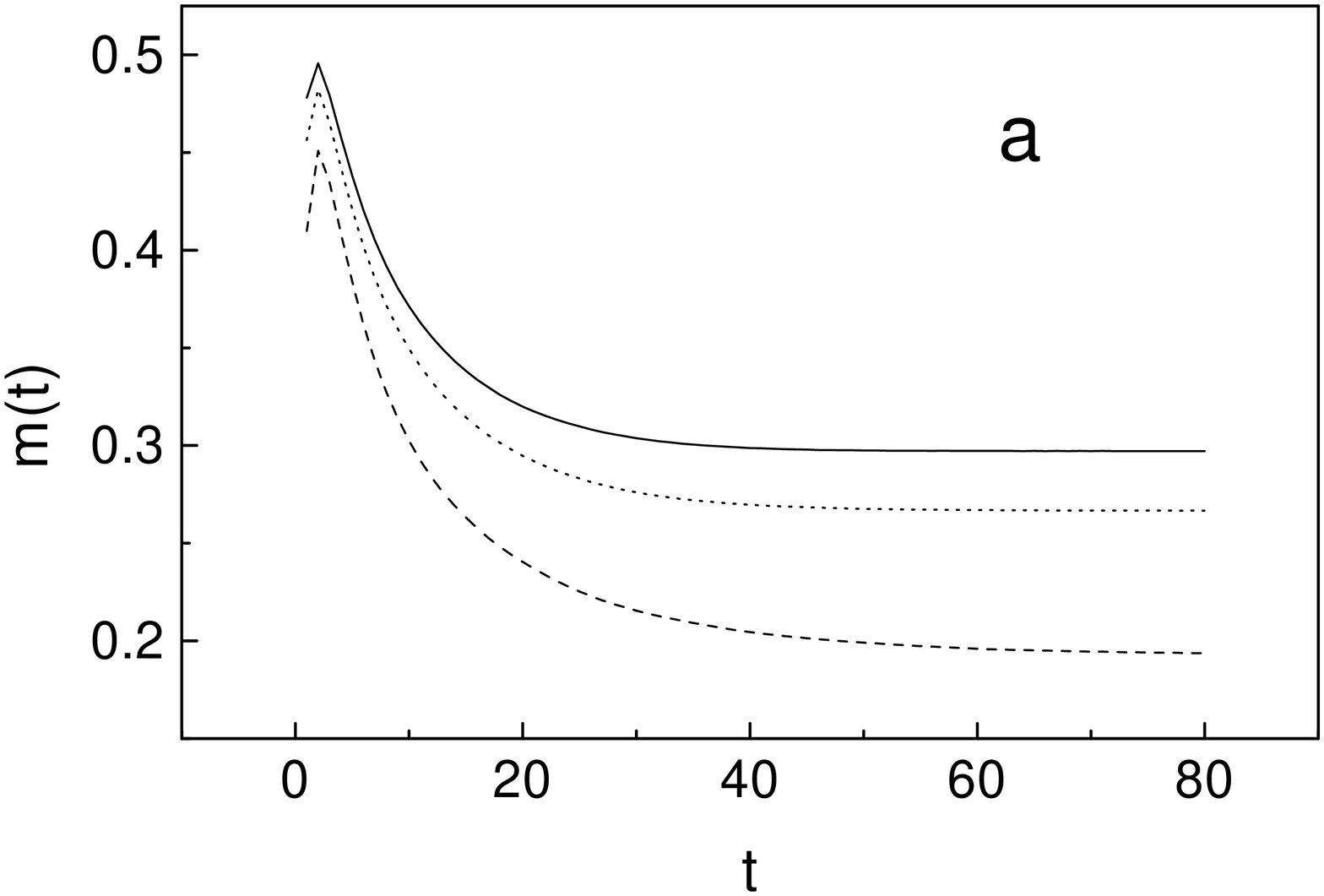}
\includegraphics[height=6.25cm,width=0.49\textwidth]{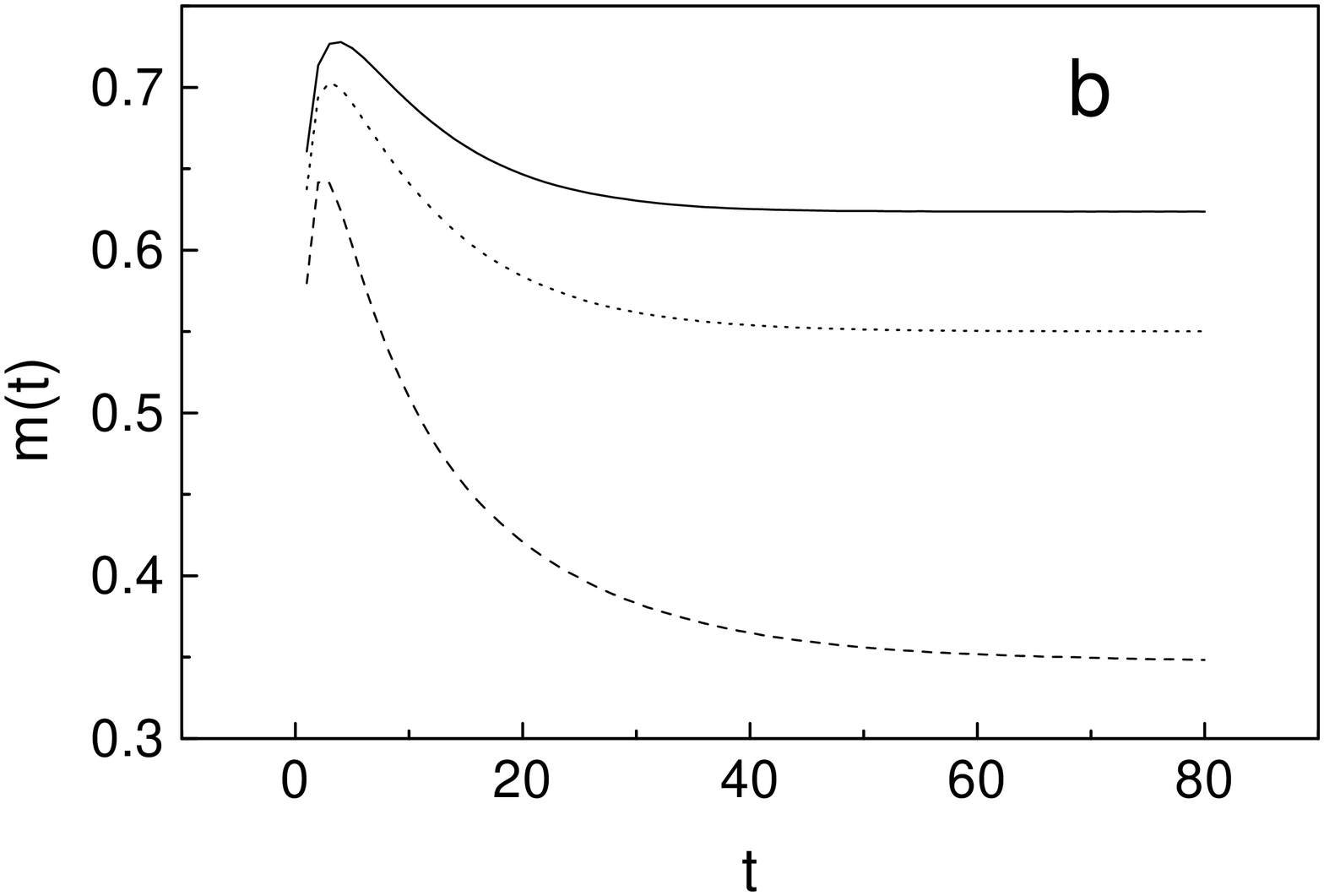}
\includegraphics[height=6.25cm,width=0.49\textwidth]{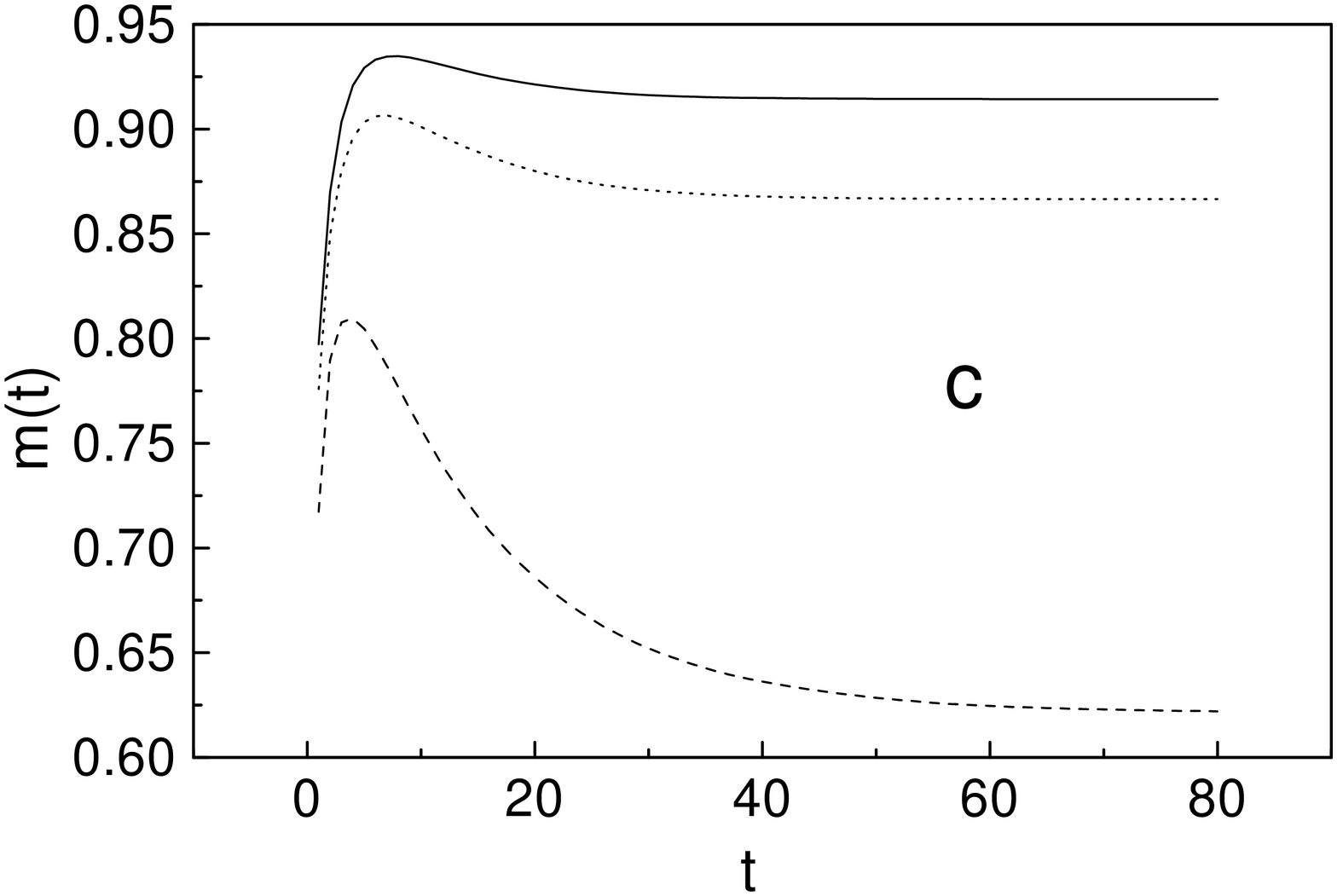}
\includegraphics[height=6.25cm,width=0.49\textwidth]{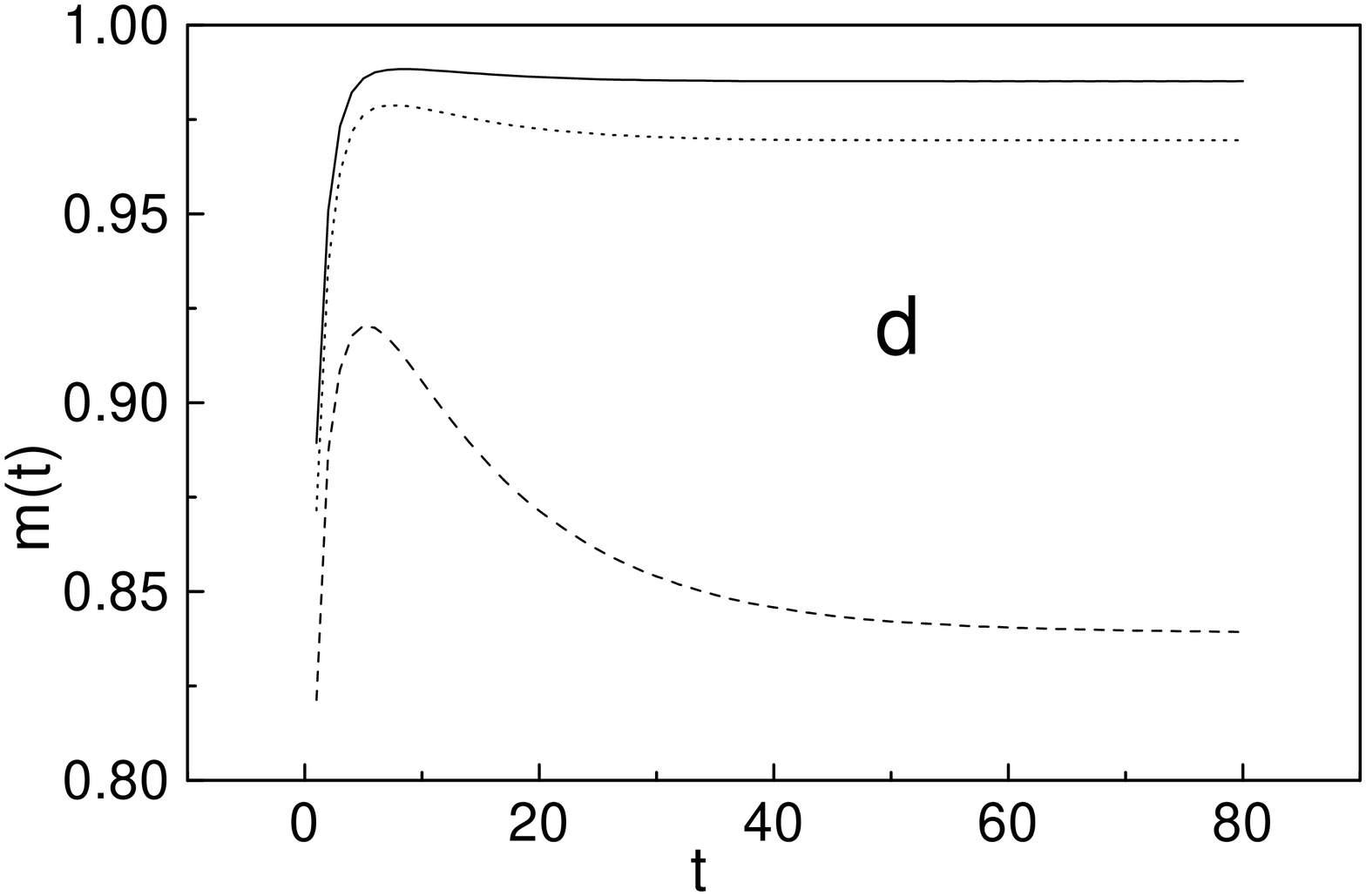}
\caption{Results of numerical simulations for the evolution of the overlap, $m_1^t$. $N=500$ and $\alpha=0.1$. Results have been averaged over 10,000 realizations of the synaptic interaction matrix for the asymmetric network and 5,000 realizations for the symmetric network. Full curve: $k=0$; dotted curve: $k=0.1$; dashed curve: $k=0.2$. (a) $m_1^0=0.2$. (b) $m_1^0=0.3$. (c) $m_1^0=0.4$ (d) $m_1^0=0.5$.}
\label{chap3_fig1}
\end{figure}

As mentioned earlier the generating functional method described above can be used to describe the dynamics of the system at any arbitrary time. However, the number of non zero order parameters grows quite fast with time and thus calculation becomes difficult. Therefore we have also carried out numerical simulation for a network of 500 spins to find out the evolution of the overlap, $m_1^t$, of the system for longer times, for various values of the initial overlap, $m_1^0$. To be specific, we have carried out simulations for $t \le 80$ to ensure that $m_1^t$ has more or less converged to its asymptotic limit. The memory loading level $\alpha$ was taken to be 0.1. The results have been averaged over 5,000 and 10,000 realizations of the synaptic interaction matrix for $k=0$ and $k\ne 0$, respectively. This provides a useful check on the analytical calculations for first two time steps. Furthermore, it also provides information about the long term behavior of the system. Results for $m_1^1$, $m_2^2$ and $m_1^{80}$ for different values of the initial overlap, $m_1^0$, and the asymmetry parameter, $k$, are given in Table \ref{chap3_table1}.
\begin{table}[htb]
\caption{Analytical results for the overlap, $m_1^t$, obtained from the generating functional technique for $t=1$ and $t=2$ for different values of the initial overlap, $m_1^0$, and the asymmetry parameter, $k$. Numbers in the parenthesis are results for the overlap for $t=1,2$ and 80, of numerical simulations of a network of 500 neurons and 50 memories. Results of simulations have been averaged over 10,000 realizations of the synaptic interaction matrix for the asymmetric network and 5,000 realizations for the symmetric network.}
\begin{center}
\begin{tabular}{|c|c|c|c|c|} \hline
$m_1^0$&$k$&$m_1^1$&$m_1^2$&$m_1^{80}$ \\ \hline
0.1&0&0.248 ($0.250 \pm 0.047$ )&0.248 ($0.247 \pm 0.078$)&($0.131 \pm 0.140$)\\
&0.1&0.237 ($0.239 \pm 0.046$)&0.243 ($0.243 \pm 0.079$)&($0.120 \pm 0.143$)\\
&0.2&0.211 ($0.211 \pm 0.045$)&0.229 ($0.229 \pm 0.079$)&($0.009 \pm 0.146$)\\ \hline
0.2&0&0.473 ($0.478 \pm 0.052$)&0.491 ($0.496 \pm 0.084$)&($0.297 \pm 0.190$)\\ 
&0.1&0.453 ($0.456 \pm 0.049$)&0.480 ($0.482 \pm 0.082$)&($0.267 \pm 0.180$)\\
&0.2&0.407 ($0.410 \pm 0.044$)&0.447 ($0.451 \pm 0.078$)&($0.194 \pm 0.160$)\\ \hline
0.3&0&0.657 ($0.661 \pm 0.051$)&0.709 ($0.713 \pm 0.083$)&($0.624 \pm 0.305$)\\
&0.1&0.634 ($0.637 \pm 0.048$)&0.690 ($0.694 \pm 0.080$)&($0.550 \pm 0.307$)\\
&0.2&0.577 ($0.580 \pm 0.043$)&0.638 ($0.642 \pm 0.075$)&($0.348 \pm 0.267$)\\ \hline
0.4&0&0.794 ($0.797 \pm 0.045$)&0.867 ($0.870 \pm 0.064$)&($0.914 \pm 0.190$)\\
&0.1&0.772 ($0.776 \pm 0.043$)&0.846 ($0.849 \pm 0.064$)&($0.867 \pm 0.235$)\\
&0.2&0.715 ($0.717 \pm 0.039$)&0.786 ($0.790 \pm 0.063$)&($0.622 \pm 0.343$)\\ \hline
0.5&0&0.886 ($0.889 \pm 0.035$)&0.950 ($0.951 \pm 0.035$)&($0.985 \pm 0.068$)\\
&0.1&0.868 ($0.871 \pm 0.034$)&0.934 ($0.936 \pm 0.040$)&($0.969 \pm 0.105$)\\
&0.2&0.818 ($0.821 \pm 0.033$)&0.883 ($0.887 \pm 0.046$)&($0.839 \pm 0.263$) \\ \hline

\end{tabular}
\end{center}
\label{chap3_table1}
\end{table}
 It can be easily seen that the results of the analytical calculation for $m_1^1$ and $m_1^2$ match quite well with those obtained by the numerical simulation. The overlap at any given time decreases with increasing $k$. This becomes more clear in Fig. \ref{chap3_fig1} 
where we show the evolution of the overlap obtained from the numerical simulation for different values of $k$ and $m_1^0$. It is also found that for the values of $m_1^0$ as high as 0.5, the overlap  increases with time initially for first few time steps. Thereafter it starts decreasing before converging to its asymptotic limit. This is consistent with the result reported in Ref. \cite{gardner87} for the symmetric network ($k$=0), that the overlap  is not always a monotonic function of time. Thus, the introduction of asymmetry in the synaptic matrix does not bring about any qualitative change in this behavior. In fact, it makes it even more pronounced (Fig. \ref{chap3_fig1}).

The effect of asymmetry on the size of the basins of attraction of the stored patterns and the convergence rate  was also investigated by the numerical simulations. In our simulations of a network of 500 neurons with 50 memories, we choose each initial state in such a way that it has a overlap $m_1^0$ with a specific memory state (the target state). The total number of initial states used for each value of $m_1^0$ was 20,000 for the asymmetric network and 10,000 for the symmetric network. The network was allowed to evolve for a maximum of 200 time steps to look for the fixed points of dynamics. Table \ref{chap3_table2} gives the probabilities for the correct retrieval (operationally defined as one in which the overlap of the fixed point  with the target memory exceeds 0.95) and the spurious retrieval, and the average convergence times for these retrievals for different values of $m_1^0$ and $k$. \begin{table}[htb]
\caption{Probabilities for the correct retrieval, $P_r$ (operationally defined as one in which the overlap of the fixed point  with the target memory exceeds 0.95) and the spurious retrieval, $P_s$, and the average convergence times for these retrievals, $\langle \tau_r \rangle$ and $\langle \tau_s \rangle$, respectively, for different values of $m_1^0$ and $k$. The results were obtained by numerical simulations of a network of 500 neurons and 50 memories. The total number of initial states used for each value of $m_1^0$ was 20,000 for the asymmetric network and 10,000 for the symmetric network. The network was allowed to evolve for a maximum of 200 time steps to look for the fixed points of the dynamics.}
\begin{center}
\begin{tabular}{|c|c|c|c|c|c|} \hline
$m_1^0$&$k$& $P_{r}$&$\langle \tau_{r} \rangle$&$P_{s}$&$\langle \tau_{s} \rangle$ \\ \hline
0.3&0&0.323&11&0.431&24\\
&0.1&0.226&12&0.499&27\\
&0.2&0.077&14&0.523&44\\ \hline
0.4&0&0.783&8&0.144&20\\
&0.1&0.682&8&0.209&23\\
&0.2&0.346&11&0.380&40\\ \hline
0.5&0&0.940&6&0.040&15\\
&0.1&0.892&6&0.068&18\\
&0.2&0.618&8&0.221&32\\ \hline
0.6&0&0.970&4&0.016&10\\
&0.1&0.943&5&0.033&12\\
&0.2&0.736&7&0.143&25 \\ \hline
\end{tabular}
\end{center}
\label{chap3_table2}
\end{table} We find that the probability of convergence to the retrieval fixed point decreases with increasing $k$ while that to the spurious fixed point increases. Thus, we do not find  increase in the size of the basins of attraction of the stored patterns when asymmetry is introduced in the synaptic interaction matrix of the Hopfield model. In fact, contrary to the expectations which arise from the calculation of fixed point attractors in Ref. \cite{us1}, we find reduction in the size of basins of attraction of the stored patterns with the increasing asymmetry in the synaptic interaction matrix. Moreover, the convergence time for the correct retrieval also increases.  However, the convergence time for the spurious fixed points increases much faster than that for the retrieval fixed point in presence of asymmetry in the synapses. This is a positive feature of the asymmetry so far as the retrieval performance of the network is concerned. It may be quite useful in discriminating the spurious fixed points.

\section{Summary and Discussions}
To summarize we have studied the synchronous dynamics of the Hopfield model of associative memory when a random antisymmetric part is added to the otherwise symmetric synaptic matrix. A generating functional technique was used to derive analytical expressions for the order parameters at $t=1$ and 2. We find that the overlap between the pattern under retrieval and the state of the network at $t=1$ is independent of the symmetry of the additional random part of the synaptic interaction matrix. The result may have bearing on the theories which use the results at $t=1$ to estimate quantities relevant to the retrieval performance of the network, namely, the size of basins of attraction of the stored patterns, the retrieval quality and the convergence rate. The symmetry of the synaptic interaction matrix becomes effective from the second time step, explicitly through a correlation function involving spin configurations at different times. This suggests that the prediction of long time behavior of the network from the dynamic behavior of the first few time steps may not be {\it always} correct. This was confirmed by the numerical simulation which shows that the difference in the long time and the short time behavior of the network becomes pronounced in presence of asymmetry in the synaptic interaction matrix. It is also found in  simulations that the size of basins of attraction of the stored patterns decreases with the asymmetry. Moreover, the convergence time for the retrieval  increases. These results are contrary to the expectations arising from the earlier studies based on the counting of fixed points. However, the convergence time for the spurious fixed points increases  faster than that for the retrieval fixed point in presence of asymmetry in the synapses. This is a positive feature of the asymmetry so far as the retrieval performance of the network is concerned. It may be quite useful in finding a way to discriminate the correct retrieval from  the spurious ones. 

Although the generating functional technique is exact in the thermodynamic limit and can be used, in principle, to calculate the properties of the network after an arbitrary number of time steps, the number of order parameters increase very quickly with time. Thus for all practical purposes the technique is used to calculate the order parameters only for first few time steps. Recently, Koyama {\it et al.} \cite{koyama99} have developed an approximate scheme which simplifies considerably the calculations of the order parameters for longer times for the synchronous dynamics of the (symmetric) Hopfield model. Henkel and Opper \cite{henkelOpper91} have formulated another approximate framework for the synchronous dynamics of neural network models of associative memory within the generating functional theory. Deriving its physical insight from  another dynamic formation known as the cavity method \cite{parisibook}, this consists of approximating the probability distribution of the local field by two Gaussian peaks to properly account for the memory effects which arise due to the spin correlations at different times. It would be quite interesting to extend the similar analyzes  to the asymmetric Hopfield model. There exists yet another possibility of formulating a method to study the dynamics at longer time scales within the very framework of generating functional method. Eisffeller and Opper \cite{opper92,opper94} have developed a mean field Monte Carlo method for simulating the parallel dynamics of the well known Sherrington-Kirkpatrick (SK) \cite{skmodel} model of a spin glass. The method uses the generating functional technique to derive a decoupled set of single spin equations which are exact in the thermodynamic limit. 
A Monte Carlo method is used to generate spin trajectories corresponding to the stochastic 
dynamics of the single spin, which in turn can be used to calculate the averaged quantities of 
interest. This technique has already been extended to study the synchronous dynamics of a one pattern model of associative memory \cite{us3}. A further generalization of the formalism to  the synchronous dynamics of the Hopfield model remains an open problem.
\label{summary}

\appendix
\section{Determination of the Vanishing Order Parameters} \label{appen1}
In this Appendix we show that a number of order parameters, which are solutions of the saddle point equations (\ref{speq1})--(\ref{speq9}), vanish. It has been shown in Ref . \cite{gardner87} that the the average of the form $\left\langle x^T\,\sigma^T\,S(x^t,\,\sigma^t)\right\rangle_{Z_{\pm}}$ vanish, where $S\left(x^t,\,\sigma^t\right)$ depends only on the times earlier than $T$.
\begin{equation}
s^{t_1\,t_2}  =  (1-g)\,\left\langle x^{t_1}\,\sigma^{t_1}\,
\sigma^{t_2-1} \right\rangle_{Z_+} + g\,\left\langle x^{t_1}\, \sigma^{t_1}\,
\sigma^{t_2-1} \right\rangle_{Z_-}\,. \nonumber
\end{equation}
We note that both of the expectation values in the above equation are of the form $\left\langle x^T\,\sigma^T\,S(x^t,\,\sigma^t)\right\rangle$ for $t_1 \ge t_2$ with $S(x^t,\,\sigma^t) = \sigma^t$ and thus
\begin{equation}
s^{t_1\,t_2} = 0 \quad \text{for} \quad t_1\ge t_2 \,.\nonumber
\end{equation}
\begin{equation}
p^{t_1\,t_2}  \underset{t_1 \le t_2}{=}  (1-g)\,\left\langle x^{t_1}\,
x^{t_2}\,\sigma^{t_1}\,\sigma^{t_2} \right\rangle_{Z_+} + 
g\,\left\langle x^{t_1}\,x^{t_2}\,\sigma^{t_1}\,\sigma^{t_2} \right\rangle_{Z_-}\,.\nonumber
\end{equation}
Here also  the expectation values are of the form $\left\langle x^T\,\sigma^T\,S(x^t,\,\sigma^t)\right\rangle$ for $t_1 < t_2$ with $S(x^t,\,\sigma^t) = \sigma^t\,x^t$ and hence
\begin{equation}
p^{t_1\,t_2}=0 \quad \text{for} \quad t_1 < t_2\,. \nonumber
\end{equation}
For $t_1=t_2=t$ we have
\begin{equation}
p^{t\,t}= (1-g)\,\left\langle \left(x^t\right)^2\right\rangle_{Z_+}
+g\,\left\langle \left(x^t\right)^2\right\rangle_{Z_-} \,.\nonumber
\end{equation}
In this case too both the expectation values are zero as we have a term 
which becomes after taking the trace over $\sigma^T$
\begin{equation}
I_T^{\prime} = \int_0^{\infty} d\lambda^T \int_{-\infty}^{\infty}\frac{dx^T}{2\,\pi}\,\left(x^T\right)^2\,\exp\left\{
i\,x^T\,\lambda^T-i\,l^{T\,T}\,\left(x^T\right)^2\right\}
\left(\mathrm{e}^{x^T\,C}-\mathrm{e}^{-x^T\,C}\right)\,.
\end{equation}
Once again by making change of variables $x^T \rightarrow -x^T$ and $\lambda^T
\rightarrow -\lambda^T$ in the second term it can be shown that 
\begin{equation}
I_T^{\prime} = 0\,.
\end{equation}
Therefore
\begin{equation}
p^{t\,t}=0\,.\nonumber
\end{equation}
Next we look at the expression for $n_1^t$
\begin{equation}
n_1^t =  (1-g)\,\left\langle \sigma^0\,x^{t+1}\,
\sigma^{t+1} \right\rangle_{Z_+} - g\,\left\langle \sigma^0\,x^{t+1}\,
\sigma^{t+1} \right\rangle_{Z_-}\,.\nonumber
\end{equation}
Here also the expectation values are of the form $\left\langle x^T\,\sigma^T\,S(x^t,\,\sigma^t)\right\rangle$  with $S(x^t,\,\sigma^t) = 
\sigma^0$ and hence
\begin{equation}
n_1^{t}=0 \quad \text{for all} \quad t\,. \nonumber
\end{equation} 
Using these results we get the following expression for $W$
\begin{eqnarray}
W &=& \int_{\infty}^{\infty} \prod_{t=1}^{T} \left[ \frac{N\,dm^{t-1}\,dn^{t-1}}{2\,\pi}\right] \exp \left\{i \,N \sum_t n^{t-1}\,
m^{t-1} - \frac{N}{2} \sum_t \left(n^{t-1}\right)^2  \right. \nonumber \\
&& \left. -N\sum_{t_1 < t_2} q^{t_1\,t_2}\,n^{t_1-1}\,n^{t_2-1}  
-N \sum_{t_1 < t_2} s^{t_1\,t_2} m^{t_1-1} n^{t_2-1} \right\}\,.
\end{eqnarray}
Let us now consider any expectation value of the form $\left\langle n^T\,n^t \right\rangle_W$ 
for $t<T$. It will have a term of the form
\begin{eqnarray}
I^{\prime \prime}_T &=& \int_{\infty}^{\infty}  \left[ \frac{N\,dm^T\,dn^T}{2\,\pi}\right]\,n^T\, \exp \left\{i \,N \, n^T\,
m^T - \frac{N}{2}\,\left(n^T\right)^2 \right. \nonumber \\
&& \left.- N\,n^T\sum_{t=0}^{T-1} q^{t+1,\,T+1}\,n^t - N\,n^T \sum_{t=0}^{T-1} s^{t+1,\,T+1} m^t \right\} \,.
\end{eqnarray}
After performing integration over $m^T$ we get
\begin{eqnarray}
I^{\prime \prime}_T &=& \int_{\infty}^{\infty} \frac{dn^T}{2\,\pi}\,\delta\left(n^T\right)\,n^T\,
\exp \left\{-\frac{N}{2}\,\left(n^T\right)^2 \right. \nonumber \\
&& \left.- N\,n^T\sum_{t=0}^{T-1} q^{t+1,\,T+1}\,n^t - N\,n^T \sum_{t=0}^{T-1} s^{t+1,\,T+1} m^t \right\}\nonumber \\
&=&0\,.
\end{eqnarray}  
Hence 
\begin{equation}
\left\langle n^{t_1-1}\,n^{t_2-1}\right\rangle_W = 0 \quad \text{for} \quad t_1 < t_2 \,.\nonumber
\end{equation}
Using similar arguments we get
\begin{equation}
\left\langle m^{t_1-1}\,n^{t_2-1}\right\rangle_W = 0 \quad \text{for} \quad t_1 < t_2 \,.\nonumber
\end{equation} 
\section{Calculation of the Non-Zero Order Parameters} \label{appen2}
In this Appendix we give the results for the order parameters and the overlap up to second time step, obtained from the saddle point equations (\ref{nonzerosp1}) -- (\ref{nonzerosp7}). 
\subsection{Order Parameters for $t=1$}
\begin{eqnarray}
i\,k^{11}&=&N\alpha\langle m^0\,n^0 \rangle_W \nonumber \\
&=& \frac{N \alpha \int_{-\infty}^{\infty} \frac{N dm^0 dn^0}{2\pi}n^0 m^0\exp\left(iNn^0m^0-\frac{N}{2}(n^0)^2\right)}{\int_{-\infty}^{\infty}\frac{N dm^0 dn^0}{2\pi}\exp\left(iNn^0m^0-\frac{N}{2}(n^0)^2\right)} \nonumber \\
&=& N \alpha \frac{1}{i}\left[ \frac{\partial}{\partial x} \ln \int_{-\infty}^{\infty}\frac{N dm^0 dn^0}{2\pi}\exp\left(ixn^0m^0-\frac{N}{2}(n^0)^2\right)\right]_{x=N} \nonumber \\
&=& N \alpha \frac{1}{i}\left[ \frac{\partial}{\partial x} \ln \int_{-\infty}^{\infty} N dn^0 \delta(x n^0) \exp\left(-\frac{N}{2}(n^0)^2\right)\right]_{x=N} \nonumber \\
&=& N \alpha \frac{1}{i}\left[ \frac{\partial}{\partial x} \ln \frac{N}{x} \right]_{x=N} \nonumber \\
\Longrightarrow k^{11}&=&\alpha \,. \nonumber
\end{eqnarray} 
\begin{eqnarray}
i\,l^{11}&=&\frac{k^2}{2} + \frac{N\alpha}{2}\left\langle (m^0)^2 \right\rangle_W \nonumber \\
\frac{N\alpha}{2}\left\langle (m^0)^2 \right\rangle_W&=& -\left[ \frac{\partial}{\partial x} \ln \int_{-\infty}^{\infty}\frac{N dm^0 dn^0}{2\pi}\exp\left(iNn^0m^0-\frac{N}{2}(n^0)^2 - \frac{N \alpha}{2} x (m^0)^2 \right)\right]_{x=0} \nonumber \\
&=&-\left[ \frac{\partial}{\partial x} \ln (\alpha x + 1)^{-1/2} \right]_{x=0} \nonumber \\
&=& \frac{\alpha}{2}\,. \nonumber 
\end{eqnarray}
\begin{eqnarray}
\Longrightarrow i l^{11}&= &\frac{1}{2} (\alpha + k^2) \,. \nonumber
\end{eqnarray}
Substituting the values of $k^{11}$ and $l^{11}$ in the expression for $Z_{\pm}$ in Eq. (\ref{zpm}) 
\begin{equation}
\langle \sigma^0 \sigma^1 \rangle_{Z_{\pm}}=\frac{\mathrm{Tr}_{\sigma^0 \sigma^1} \, \sigma^0 \sigma^1 \int_0^{\infty} d\lambda^1 \int_{-\infty}^{\infty} \frac{dx^1}{2\pi}\exp \left\{ i x^1 \lambda^1 -\frac{1}{2}(\alpha + k^2)(x^1)^2 \mp i \sigma^0 m^0_1 x^1 \sigma^1 \right\}}{\mathrm{Tr}_{\sigma^0 \sigma^1} \int_0^{\infty} d\lambda^1 \int_{-\infty}^{\infty} \frac{dx^1}{2\pi}\exp \left\{ i x^1 \lambda^1 -\frac{1}{2}(\alpha + k^2)(x^1)^2 \mp i \sigma^0 m^0_1 x^1 \sigma^1 \right\}} \,.\nonumber
\end{equation}
Performing the trace we get
\begin{equation}
\langle \sigma^0 \sigma^1 \rangle_{Z_{\pm}}=\frac{\int_0^{\infty} d\lambda^1 \int_{-\infty}^{\infty} \frac{dx^1}{2\pi}\mathrm{e}^{ i x^1 \lambda^1 -\frac{1}{2}(\alpha + k^2)(x^1)^2}\left[\mathrm{e}^{\mp i m^0_1 x^1} - \mathrm{e}^{\pm i m^0_1 x^1} \right]}{\int_0^{\infty} d\lambda^1 \int_{-\infty}^{\infty} \frac{dx^1}{2\pi}\mathrm{e}^{ i x^1 \lambda^1 -\frac{1}{2}(\alpha + k^2)(x^1)^2}\left[\mathrm{e}^{\mp i m^0_1 x^1} + \mathrm{e}^{\pm i m^0_1 x^1} \right]}\,.\nonumber
\end{equation}
For the second term in the denominator let $x^1\rightarrow -x^1$ and $\lambda \rightarrow -\lambda$ so that
\begin{eqnarray}
\langle \sigma^0 \sigma^1 \rangle_{Z_{\pm}}&=&\frac{\int_0^{\infty} \frac{d\lambda^1}{2\pi(\alpha+k^2)} \left[\mathrm{e}^{-(\lambda^1 \mp  m^0_1)^2/2(\alpha+k^2)} - \mathrm{e}^{-(\lambda^1 \pm  m^0_1)^2/2(\alpha+k^2)}  \right]}{\int_{-\infty}^{\infty} d\lambda^1 \int_{-\infty}^{\infty} \frac{dx^1}{2\pi}\mathrm{e}^{ i x^1 \lambda^1 -\frac{1}{2}(\alpha + k^2)(x^1)^2 \mp i m^0_1 x^1}} \nonumber \\
&=&\frac{\frac{1}{2}\left[ \mathrm{erfc}\left(\mp \frac{m_1^0}{\sqrt{2(\alpha+k^2)}}\right)- \mathrm{erfc}\left(\pm \frac{m_1^0}{\sqrt{2(\alpha+k^2)}}\right) \right]}{\int_{-\infty}^{\infty} dx^1 \delta(x)\mathrm{e}^{-\frac{1}{2}(\alpha + k^2)(x^1)^2 \mp i m^0_1 x^1}} \nonumber \\
&=&\pm \mathrm{erf}\left(\pm \frac{m_1^0}{\sqrt{2(\alpha+k^2)}}\right)\nonumber \\
m^1_1&=&(1-g)\langle \sigma^0 \sigma^1 \rangle_{z_+} - g \langle \sigma^0 \sigma^1 \rangle_{z_-} \nonumber \\
&=& \mathrm{erf}\left(\frac{m_1^0}{\sqrt{2(\alpha+k^2)}}\right)\,. \nonumber
\end{eqnarray} 
\subsection{Order Parameters for $t=2$}
\begin{eqnarray}
q^{12}&=&(1-g)\langle \sigma^0 \sigma^1 \rangle_{z_+} + g \langle \sigma^0 \sigma^1 \rangle_{z_-} \nonumber \\
&=&(1-2g)m^1_1 = m_1^0 m^1_1 \,. \nonumber
\end{eqnarray}
\begin{eqnarray}
s^{12}&=&(1-g)\langle x^1 \rangle_{Z_+}+g \langle x^1 \rangle_{Z_-} \nonumber
\end{eqnarray}
In the expressions for the expectation values we have an integral of the form which comes from $Z_{\pm}$:
\begin{eqnarray}
I&=&\mathrm{Tr}_{\sigma^2} \int_0^{\infty} d\lambda^2 \int_{-\infty}^{\infty} \frac{dx^2}{2\pi} \mathrm{e}^{i\lambda^2 x^2 - i l^{22}(x^2)^2 + x^2 \sigma^2(i\alpha-i k^{21}\sigma^0 - i k^{22} \sigma^1 - i l^{12} x^1 \sigma^1 \mp i \sigma^0 m_1^1)} \nonumber
\end{eqnarray}
First performing the trace over $\sigma^2$ and making $x^2 \rightarrow -x^2$ and $\lambda^2 \rightarrow -\lambda^2$ in the second term allows us to combine both the terms with limit of integration over $\lambda^2$ from 0 to $\infty$ to $-\infty$ to $\infty$. Thus, we have  
\begin{eqnarray}
I&=&\int_{-\infty}^{\infty} d\lambda^2 \int_{-\infty}^{\infty} \frac{dx^2}{2\pi} \mathrm{e}^{i\lambda^2 x^2 - i l^{22}(x^2)^2 + x^2 (i\alpha-i k^{21}\sigma^0 - i k^{22} \sigma^1 - i l^{12} x^1 \sigma^1 \mp i \sigma^0 m_1^1)} \nonumber \\
&=&\int_{-\infty}^{\infty} dx^2 \delta(x^2) \mathrm{e}^{- i l^{22}(x^2)^2 + x^2 (i\alpha-i k^{21}\sigma^0 - i k^{22} \sigma^1 - i l^{12} x^1 \sigma^1 \mp i \sigma^0 m_1^1)} =1 \nonumber 
\end{eqnarray}
\begin{eqnarray}
\Longrightarrow \langle x^1 \rangle_{Z_{\pm}}&=&\frac{\frac{1}{2}\mathrm{Tr}_{\sigma^0 \sigma^1} \int_0^{\infty} d\lambda^1 \int_{-\infty}^{\infty} \frac{dx^1}{2\pi} x^1  \,\mathrm{e}^{i \lambda^1 x^1 - i l^{11} (x^1)^2 \mp i \sigma^0 \sigma^1 m_1^0 x^1}} {\frac{1}{2}\mathrm{Tr}_{\sigma^0 \sigma^1} \int_0^{\infty} d\lambda^1 \int_{-\infty}^{\infty} \frac{dx^1}{2\pi} \, \mathrm{e}^{i \lambda^1 x^1 - i l^{11} (x^1)^2 \mp i \sigma^0 \sigma^1 m_1^0 x^1}} \nonumber \\
&=&\int_0^{\infty} d\lambda^1 \int_{-\infty}^{\infty} \frac{dx^1}{2\pi} x^1  \,\mathrm{e}^{i \lambda^1 x^1 - (\alpha+k^2) (x^1)^2/2} \left[\mathrm{e}^{ \mp i m_1^0 x^1}+\mathrm{e}^{ \pm i m_1^0 x^1}\right] \nonumber \\
&=&i \sqrt{\frac{2}{\pi (\alpha+k^2)}} \exp \left\{-\frac{(m_1^0)^2}{2(\alpha+k^2)}\right\} \nonumber \\
%\end{eqnarray}
%\begin{eqnarray}
\Longrightarrow s^{12}&=&i \sqrt{\frac{2}{\pi (\alpha+k^2)}}\exp \left\{-\frac{(m_1^0)^2}{2(\alpha+k^2)}\right\} \,. \nonumber 
\end{eqnarray}
\begin{eqnarray}
&i k^{22}&=N \alpha \langle m^1 n^1 \rangle_W \nonumber \\
&=&N \alpha \frac{\int_{-\infty}^{\infty}\frac{N dm^0 dn^0}{2\pi}\frac{N dm^1 dn^1}{2\pi} m^1 n^1 \mathrm{e}^{iN n^0 m^0 + i N m^1 n^1 -\frac{N}{2}(n^0)^2 -\frac{N}{2}(n^1)^2 - N q^{12} n^0 n^1 - N s^{12} m^0 n^1}} {\int_{-\infty}^{\infty}\frac{N dm^0 dn^0}{2\pi}\frac{N dm^1 dn^1}{2\pi}  \mathrm{e}^{iN n^0 m^0 + i N m^1 n^1 -\frac{N}{2}(n^0)^2 -\frac{N}{2}(n^1)^2 - N q^{12} n^0 n^1 - N s^{12} m^0 n^1}} \nonumber 
\end{eqnarray}
it can easily seen that the integration over $m^0$ gives $\delta (n^0 + i s^{12} n^1)$ so that  \begin{eqnarray}
&i k^{22}&=N \alpha \frac{\int_{-\infty}^{\infty}\frac{N dm^1 dn^1}{2\pi} m^1 n^1 \mathrm{e}^{i N m^1 n^1 +\frac{N}{2}(s^{12})^2 (n^1)^2-\frac{N}{2}(n^1)^2 + i N q^{12} s^{12} (n^1)^2 }} {\int_{-\infty}^{\infty}\frac{N dm^1 dn^1}{2\pi} \mathrm{e}^{i N m^1 n^1 +\frac{N}{2}(s^{12})^2 (n^1)^2-\frac{N}{2}(n^1)^2 + i N q^{12} s^{12} (n^1)^2 }} \nonumber \\
&=&\frac{N \alpha}{i}\left[ \frac{\partial}{\partial x} \ln \int_{-\infty}^{\infty}\frac{N dm^1 dn^1}{2\pi} \mathrm{e}^{i x m^1 n^1 +\frac{N}{2}(s^{12})^2 (n^1)^2-\frac{N}{2}(n^1)^2 + i N q^{12} s^{12} (n^1)^2 }\right]_{x=N} \nonumber  
\end{eqnarray}
Integration over $m^1$ gives $\delta(x n^1)$ and we get
\begin{eqnarray}
i k^{22}&=&\frac{N \alpha}{i}\left[ \frac{\partial}{\partial x} \ln \frac{N}{x}\right]_{x=N}=i \alpha \nonumber \\
\Longrightarrow k^{22}&=&\alpha \,. \nonumber 
\end{eqnarray}
Now 
\begin{equation}
i k^{21}=-k^2 s^{12}  + N \alpha \langle m^1 n^0 \rangle_{W} \nonumber
\end{equation}
Proceeding in the same way as in the calculation of $k^{22}$ given above we note that the presence of  $\delta (n^0 + i s^{12} n^1)$ (which appears after performing integration over $m^0$) can be exploited to have
\begin{equation}
N \alpha \langle m^1 n^0 \rangle_{W}=-i s^{12} N \alpha \langle m^1 n^1 \rangle_{W}= \alpha s^{12} \nonumber
\end{equation}
Therefore
\begin{eqnarray}
i k^{21}&=&(\alpha - k^2) s^{12} \nonumber \\
\text{or} \quad k^{21}&=&(\alpha-k^2) \sqrt{\frac{2}{\pi (\alpha+k^2)}}\exp \left\{-\frac{(m_1^0)^2}{2(\alpha+k^2)}\right\} \,. \nonumber
\end{eqnarray}
In order to calculate $l^{22}$ we need to evaluate the following expectation value

\begin{eqnarray}
\frac{N \alpha}{2} \langle (m^1)^2 \rangle_{W}&=&-\left[\frac{\partial}{\partial h} \ln   \int_{-\infty}^{\infty}\frac{N dm^1 dn^1}{2\pi} \mathrm{e}^{ -\frac{N}{2}(n^1)^2\{1-2 i q^{12} s^{12} -(s^{12})^2\} -i N m^1 n^1 -\frac{N \alpha}{2} h (m^1)^2 } \right]_{h=0} \nonumber \\
&=& -\left[\frac{\partial}{\partial h}  \ln\left[ \{1-2 i q^{12} s^{12} -(s^{12})^2 \} \alpha h +1 \right]^{-1/2}\right]_{h=0} \nonumber \\
&=&\frac{\alpha}{2}\{1-2 i q^{12} s^{12} -(s^{12})^2 \} \nonumber
\end{eqnarray}
Here, too, we have used the result that the integration over $m^0$ gives $\delta (n^0+i s^{12} n^1)$. Now
\begin{eqnarray}
i l^{22}&=&\frac{k^2}{2} + \frac{N \alpha}{2} \langle (m^1)^2 \rangle_{W}=\frac{1}{2}(\alpha+k^2) -\frac{\alpha}{2}\{(s^{12})^2 +2 i q^{12} s^{12} \}\nonumber
\end{eqnarray}
Calculation of $l^{12}$ requires 
\begin{eqnarray}
N \alpha \langle m^0 m^1 \rangle_{W}&=&-i \left[\frac{\partial}{\partial x} \ln  \int_{-\infty}^{\infty}\frac{N dm^0 dn^0}{2\pi} \int_{-\infty}^{\infty}\frac{N dm^1 dn^1}{2\pi}\,\mathrm{e}^{i N m^0 n^0 + i N m^1 n^1 + i x N \alpha m^0 m^1 } \right. \nonumber \\
&&\left. \times \mathrm{e}^{-\frac{N}{2} (n^0)^2 -\frac{N}{2} (n^1)^2-N q^{12} n^0 n^1 - N s^{12} m^0 n^1} \right]_{x=0} \,. \nonumber
\end{eqnarray}
Integration over $m^1$ gives $\delta (n^1 + x \alpha m^0)$ and hence
\begin{eqnarray}
N \alpha \langle m^0 m^1 \rangle_{W}&=&-i \left[\frac{\partial}{\partial x} \ln  \int_{-\infty}^{\infty}\frac{N dm^0 dn^0}{2\pi} \,\mathrm{e}^{i N m^0 n^0  -\frac{N}{2} (n^0)^2 -\frac{N}{2} x^2 \alpha^2 (m^0)^2 + N q^{12} x \alpha n^0 m^0 + N s^{12} x \alpha (m^0)^2 } \right]_{x=0}  \nonumber \\
&=&-i\left[ \frac{\partial}{\partial x} \ln  \{ x^2 \alpha^2 -2 x \alpha s^{12}  + (1 - i q^{12} x \alpha)^2 \}^{-1/2} \right]_{x=0}  \nonumber \\
&=&\alpha q^{12} - i \alpha s^{12} \nonumber 
\end{eqnarray}
Therefore
\begin{equation}
i l^{12}=k^2 q^{12}+N\alpha \langle m^0 m^1 \rangle_{W}=(k^2+\alpha) q^{12} - i \alpha s^{12}\,. \nonumber
\end{equation}
Finally we need to calculate the expectation values $\langle \sigma^0 \sigma^2 \rangle_{Z_{\pm}} $ for $m_1^2$.
\begin{eqnarray}
\langle \sigma^0 \sigma^2 \rangle_{Z_{\pm}}&=&\frac{1}{2} \mathrm{Tr}_{\sigma^0 \sigma^1 \sigma^2} \int_0^{\infty}  d\lambda^1 d\lambda^2 \int_{-\infty}^{\infty} \frac{dx^1}{2\pi} \frac{dx^2}{2\pi} \, \sigma^0 \sigma^2 \,\, \exp \left\{i \lambda^1 x^1 + i \lambda^2 x^2 - i l^{11} (x^1)^2 -  \right. \nonumber \\
&& \left. - i l^{22} (x^2)^2 - i k^{21} x^2 \sigma^2 \sigma^0 -i l^{12} x^1 x^2 \sigma^1 \sigma^2 \mp i \sigma^0 m_1^0 x^1 \sigma^1  \mp i \sigma^0 m_1^1 x^2 \sigma^2 \right\} \,. \nonumber
\end{eqnarray}
The above expression has an integral
\begin{equation}
I=\mathrm{Tr}_{\sigma^1} \int_0^{\infty} d\lambda^1 \int_{-\infty}^{\infty} \frac{dx^1}{2\pi}\,\exp \{i \lambda^1 x^1 - i l^{11} (x^1)^2 + x^1 \sigma^1 (i l^{12} x^2 \sigma^2 \mp i \sigma^0 m_1^0)\} \,.\nonumber
\end{equation}
Taking trace on $\sigma^1$ and letting $\lambda^1 \rightarrow -\lambda^1$ and $x^1 \rightarrow -x^1$ in the second term, as before, it is easy to show that $I = 1$. Thus,
\begin{eqnarray}
\langle \sigma^0 \sigma^2 \rangle_{Z_{\pm}}&=&\mathrm{Tr}_{\sigma^0 \sigma^2}\,\, \sigma^0 \sigma^2 \int_0^{\infty} d\lambda^2 \int_{-\infty}^{\infty} \frac{dx^2}{2\pi}\,\exp \{i \lambda^2 x^2 - i l^{22} (x^2)^2 -i k^{21} x^2 \sigma^2 \sigma^0 \mp i \sigma^0 m_1^1 x^2 \sigma^2 \} \nonumber \\
&=&\frac{1}{2} \int_0^{\infty} d\lambda^2 \int_{-\infty}^{\infty} \frac{dx^2}{2\pi}\,\exp \{i \lambda^2 x^2 - i l^{22} (x^2)^2 \}\left[\mathrm{e}^{-x^2(i k^{21} \pm i m_1^1)}-\mathrm{e}^{x^2(i k^{21} \pm i m_1^1)} \right] \nonumber \\
&=&\frac{1}{2}\mathrm{erfc}\left( \frac{-k^{21} \mp m_1^1}{2 \sqrt{i l^{22}}}\right)-\mathrm{erfc}\left( \frac{k^{21} \pm m_1^1}{2 \sqrt{i l^{22}}}\right) \nonumber \\
&=&\pm \mathrm{erf}\left( \frac{ m_1^1  \pm k^{21}}{2 \sqrt{i l^{22}}}\right) \,. \nonumber
\end{eqnarray}
Therefore
\begin{equation}
m_1^2=(1-g)\langle \sigma^0 \sigma^2 \rangle_{Z_+} - g \langle \sigma^0 \sigma^2 \rangle_{Z_-}=\frac{1+m_1^0}{2}\,\mathrm{erf}\left( \frac{ m_1^1  + k^{21}}{2 \sqrt{i l^{22}}}\right)+\frac{1-m_1^0}{2}\,\mathrm{erf}\left( \frac{ m_1^1  - k^{21}}{2 \sqrt{i l^{22}}}\right)\,. \nonumber
\end{equation}

\end{document}